\documentclass[aps,prd,reprint,twocolumn,10pt,eqsecnum,showpacs,nofootinbib,amsfonts,amssymb,amsmath,longbibliography,superscriptaddress,floatfix]{revtex4-1}
\usepackage[colorlinks=true,allcolors=blue]{hyperref}
\usepackage{bm}
\usepackage{mathrsfs}
\usepackage{xcolor}
\usepackage{mathtools}
\usepackage{graphicx}
\usepackage{orcidlink}

\usepackage[T1]{fontenc}

\newcommand{\ud}{\mathrm{d}}
\newcommand{\mc}[1]{\mathcal{#1}}
\newcommand{\ms}[1]{\mathscr{#1}}
\renewcommand{\Re}{\operatorname{Re}}
\renewcommand{\Im}{\operatorname{Im}}
\newcommand{\interchange}[2]{(#1 \longleftrightarrow #2)}
\newcommand{\ord}[2]{\underset{^{(#1)}}{#2}{}}

\usepackage[normalem]{ulem}

\begin{document}

\author{Alexander M.\ Grant
  \orcidlink{0000-0001-5867-4372}}
\email{a.m.grant@soton.ac.uk}
\affiliation{Department of Physics, University of Virginia, P.O.~Box 400714,
  Charlottesville, Virginia 22904-4714, USA}
\affiliation{School of Mathematical Sciences, University of Southampton,
  Southampton, SO17 1BJ, United Kingdom}

\author{Keefe Mitman
	\orcidlink{0000-0003-0276-3856}}
\email{kmitman@caltech.edu}
\affiliation{Theoretical Astrophysics 350-17, California Institute of Technology,
  Pasadena, California 91125, USA}

\title{Higher Memory Effects in Numerical Simulations of Binary Black Hole Mergers}

\begin{abstract}
	Gravitational memory effects are predictions of general relativity that are characterized by an observable effect that persists after the passage of gravitational waves.
  In recent years, they have garnered particular interest, both due to their connection to asymptotic symmetries and soft theorems and because their observation would serve as a unique test of the nonlinear nature of general relativity.
  Apart from the more commonly known displacement and spin memories, however, there are other memory effects predicted by Einstein's equations that are associated with more subleading terms in the asymptotic expansion of the Bondi-Sachs metric.
  In this paper, we write explicit expressions for these higher memory effects in terms of their charge and flux contributions.
  Further, by using a numerical relativity simulation of a binary black hole merger, we compute the magnitude and morphology of these terms and compare them to those of the displacement and spin memory.
  We find that, although these terms are interesting from a theoretical perspective, due to their small magnitude they will be particularly challenging to observe with current and future detectors.
\end{abstract}

\maketitle

\tableofcontents

\section{Introduction}

The gravitational wave memory effect (now called ``displacement memory'' to avoid confusion with other, similar effects) arises as a permanent change in the separation of two observers due to the passage of gravitational waves.
This effect was first postulated in the linearized regime~\cite{Zeldovich1974}, but for bound systems, the predominant effect is sourced by nonlinear effects in the propagation of gravitational waves~\cite{Christodoulou1991, Thorne1992}.
This effect is difficult to detect both because of its low-frequency nature and that it is simply smaller in amplitude than the predominantly oscillatory part of the gravitational wave signal.
However, there have been numerous proposals to measure it using both ground-based detectors (by stacking multiple signals in current detectors~\cite{Lasky2016, Boersma2020, Grant2022c}, which has been performed explicitly in Refs.~\cite{Hubner2019, Hubner2021}) or from single events in future detectors~\cite{Johnson2018, Grant2022c}.
It should also be seen by space-based detectors~\cite{Islo2019, Gasparotto2023, Ghosh2023} and pulsar timing arrays, and the latter currently provide upper bounds on the memory coming from the stochastic background~\cite{NANOGrav2023}.

In addition to the displacement memory, a variety of other effects have been considered in the literature which appear as permanent changes in some idealized detector (for a summary, see~\cite{Flanagan2019a}).
While this paper does not focus on the observable nature of these effects, the ``curve deviation'' observable of~\cite{Flanagan2019a} provides a motivation for studying the quantities of interest in this paper as a simple generalization of the displacement memory effect.
As shown in~\cite{Grant2021b}, the curve deviation near null infinity can be written in terms of the temporal moments\footnote{These temporal moments can be related to the integer values of the Mellin transform: see the discussion in Sec.~\ref{sec:moments} below.} of the Bondi news, a tensor which characterizes the presence of radiation at null infinity.
Much like the moments of a probability distribution, these moments characterize the Bondi news as a function of time.
The zeroth moment (that is, a single integral) of the news is exactly the displacement memory effect, but with each higher moment of the news one can recover additional observables.
For example, the spin~\cite{Pasterski2015, Nichols2017} and center-of-mass~\cite{Nichols2018} memories form the electric- and magnetic-parity pieces of the ``drift memory''\footnote{Note that this effect was previous called the ``subleading displacement memory'' in~\cite{Flanagan2019a}; see Ref.~\cite{Grant2024} for an explanation for the change in terminology.}, which is related to the first moment of the news.
Collectively, we refer to the infinite collection of moments of the news as ``higher memory effects''.

More relevantly for this paper, Ref.~\cite{Grant2021b} showed that these moments of the news could be written in terms of a ``charge'' contribution and a ``flux'' contribution, which generalizes the usual linear/nonlinear or ordinary/null~\cite{Bieri2013b} splittings of the displacement memory.
For the drift memory, such a decomposition was already known~\cite{Pasterski2015, Nichols2018}, but that such a splitting could be achieved for \emph{all} higher memory effects was a novel result, and follows from the asymptotic form of the Einstein equations.
We review this construction explicitly for the zeroth, first, and second moments in Sec.~\ref{sec:moment_decomp} below.

Asymptotic symmetries provide another formulation of the displacement memory that is frequently discussed in the literature (see, for example,~\cite{Strominger2017} and the references therein).
Here, for each spherical harmonic mode, the charge contribution can be thought of as the change in a charge constructed from a member of the supertranslation subalgebra of the Bondi-Metzner-Sachs (BMS) algebra---the symmetry algebra at null infinity. Understanding the drift memory in this way, however, requires an extension of the BMS algebra into either of the ``extended''~\cite{Barnich2009} or ``generalized''~\cite{Campiglia2014} BMS algebras, and many issues with these extensions have been pointed out in the literature~\cite{Flanagan2015, Flanagan2019b, Elhashash2021}.
Recently, an infinite set of charges were also defined which form a representation of the algebra $Lw_{1 + \infty}$~\cite{Freidel2021}, and at linear order these charges can be shown~\cite{Compere2022} to be equivalent to those in Ref.~\cite{Grant2021b}.
While these connections to symmetry algebras are interesting, we do not discuss them further in this paper, as everything that we present can be straightforwardly derived from the Einstein equations.

An important question remains about the status of this split into charge and flux contributions: for a given system, which is larger?
This is important for observing the memory effect, by the following argument~\cite{Grant2022c}.
For simplicity, we restrict to a discussion of the displacement memory, although similar arguments for higher memories also apply.
First, note that by ``detecting the displacement memory'', one does not mean ``measuring some finite offset in the detector'', as current ground-based detectors are only sensitive in some frequency range that does not include zero.
Instead, one should try to detect a part of the signal which contributes to this finite offset, which one could call the time-dependent ``memory signal''.
In the case of the displacement memory, it happens to be the case that, for binary mergers, it is the flux contribution which makes the largest contribution (for the $m = 0$ modes) to the zeroth moment of the news (see, for example, Ref.~\cite{Mitman2020}).
Moreover, since this contribution as a function of time is non-oscillatory and looks like a smoothed out step function, it is a reasonable choice for the time-dependent ``displacement memory signal''.

When considering higher memories, one is faced with another issue: current ground-based gravitational wave detectors are sensitive to the \emph{shear}, and not any particular integrals of the news.
If one is given the $n$th moment of the news as a function of time, one can obtain the shear as a function of time (up to an arbitrary initial value) by differentiating $n$ times.\footnote{See Eq.~\eqref{eq:momentofnews} for the explicit relationship between the shear and the moments of the news.}
As such, the question of which is the larger contribution should be asked of the $n$th derivative of the charge or flux.
For the case of the spin memory (the magnetic part of the first moment of the news), it is once again the flux contribution that is larger. Furthermore, this contribution is again non-oscillatory~\cite{Mitman2020} (in the $m = 0$ modes) and resembles a single bump, i.e., a smoothed-out delta function.

Motivated by these considerations, we seek to answer the following questions for the higher memory effects:
\begin{enumerate}

\item Which quantity contributes more significantly to the shear: charge or flux?

\item Is it always the case that the charge contributions are primarily oscillatory, and the flux contributions are primarily non-oscillatory?

\item What is the general morphology of these non-oscillatory contributions?

\end{enumerate}
As there are an infinite number of these higher memory effects, we restrict our attention to those associated with the zeroth, first, and second moments of the news: the displacement, drift, and ``ballistic''~\cite{Grant2024} memory effects.
To explore these questions, we will examine the waveforms produced by numerical relativity simulations of binary black hole mergers that have been run using the Simulating eXtreme Spacetimes (SXS) Collaboration's SpEC code (to perform the Cauchy evolution) and SpECTRE code [to extract the asymptotic data via a Cauchy-characteristic evolution (CCE)]~\cite{spectrecode, Moxon2020, Moxon:2021gbv}.
The code which we used to produce these waveforms is publicly available at~\cite{HigherMemoryEffects}.

The structure of this paper is the following. In Sec.~\ref{sec:background}, we review the necessary material for this paper: Bondi-Sachs coordinates, the tetrad variables that we will work with (the shear $\sigma$ and modified Weyl scalars $\psi_i$), and the moments of the news. In Sec.~\ref{sec:decomposition}, we then show how one can (non-uniquely) define charges whose evolution gives these moments of the news, adapting the procedure in Ref.~\cite{Grant2021b} to be in terms of tetrad variables, instead of tensorial quantities.
We also discuss the procedure for inverting the moments of the news in order to recover the shear.
Finally, in Sec.~\ref{sec:results}, we discuss our numerical results for binary black hole mergers, showing the hierarchy of the various contributions to the shear, as well as their morphology.
We present our conclusions in Sec.~\ref{sec:discussion}.

In this paper, we set $G=c=1$ and adopt the mostly plus metric signature.
We use Latin characters from the beginning of the alphabet ($a$, $b$, etc.) for abstract indices and Latin characters from the middle of the alphabet ($i$, $j$, etc.) in order to denote indices associated with coordinates on the two-sphere (for example, spherical coordinates $\{\theta, \phi\}$).

\section{Background} \label{sec:background}

\subsection{Bondi-Sachs coordinates}

In this paper, we work with asymptotic quantities defined with respect to a tetrad, and using the Geroch-Held-Penrose (GHP) formalism~\cite{Geroch1973}.
For concreteness, though, we begin with the form of the metric in Bondi-Sachs coordinates, which we denote with $u$, $r$, and two arbitrary angular coordinates $\theta^i$.
In vacuum general relativity, this metric takes the form (see, for example,~\cite{Grant2021b})
\begin{widetext}
\begin{equation} \label{eqn:bondi}
  \ud s^2 = -\left(1 - \frac{2V}{r}\right) e^{2 \beta/r^2} \ud u^2 - 2 e^{2 \beta/r^2} \ud u \ud r + r^2 \left(\gamma h_{ij} + \frac{1}{r} \mc C_{ij}\right) \left(\ud \theta^i - \frac{\mc U^i}{r^2} \ud u\right) \left(\ud \theta^j - \frac{\mc U^j}{r^2} \ud u\right),
\end{equation}
\end{widetext}
where $h_{ij}$ is the metric on the two-sphere (which we use to raise and lower $\theta^i$ indices).
The scalar $\gamma$ is given by
\begin{equation}
  \gamma = \sqrt{1 + \frac{\mc C_{ij} \mc C^{ij}}{2r^2}},
\end{equation}
which enforces the Bondi condition that the determinant of the angular part of the metric is constant in $r$.
There are six remaining metric functions that arise through $V$, $\beta$, $\mc U^i$, and $\mc C_{ij}$ (which is trace-free with respect to $h_{ij}$).
For this paper, $\beta$ is unimportant, but the remaining five have the following expansions:
\begin{subequations}
  \begin{align}
    V &= m + \mc M/r, \\
    \mc C_{ij} &= C_{ij} + \frac{1}{r^2} \mc E_{ij}, \\
    \mc U^i &= -\frac{1}{2} \ms D_j C^{ij} - \frac{1}{r} \bigg[\frac{2}{3} N^i - \frac{1}{16} \ms D^i (C_{jk} C^{jk}) \nonumber \\
    &\hspace{8em}- \frac{1}{2} C^{ij} \ms D^k C_{jk}\bigg] + \frac{1}{r^2} \Upsilon^i,
  \end{align}
\end{subequations}
where $\ms D_i$ is the derivative on the sphere compatible with $h_{ij}$.
Note that, in terms of their $r$ dependence, $\mc M$, $\mc E_{ij}$, and $\Upsilon^i$ are all $O(1)$ functions of $r$, but $m$, $C_{ij}$, and $N^i$ are all constants.
Again, $\mc M$ and $\Upsilon^i$ are unimportant for the discussion in this paper, but we write $\mc E_{ij}$ explicitly as a power series in $r$:
\begin{equation}
  \mc E_{ij} = \sum_{n = 0}^\infty \frac{1}{r^n} \ord{n}{\mc E}_{ij}.
\end{equation}

As a consequence of Einstein's equations, the free data at null infinity in vacuum general relativity are the initial values of $m$, $N^i$, and each $\ord{n}{\mc E}_{ij}$, together with the value of $C_{ij}$ at all times.
Due to their importance, these quantities all have names:
\begin{itemize}

\item $m$ is the \emph{mass aspect}, a generalization of the mass that is angle-dependent;

\item $N^i$ is the \emph{angular momentum aspect}, with a similarly intuitive explanation for its name;

\item $C_{ij}$ is the \emph{shear}, since (as we discuss later, in Sec.~\ref{sec:ghp}) it is related to the shear in the GHP formalism; and

\item $\ord{n}{\mc E}_{ij}$ are the ``higher Bondi aspects''~\cite{Compere2022}.

\end{itemize}
Moreover, the shear $C_{ij}$, being the $1/r$, traceless and transverse part of the metric, is related to the usual transverse-traceless metric that an observer will measure at large distances from a source.
Finally, a quantity of great interest in these spacetimes is the \emph{news} $N_{ij}$:
\begin{equation}
  N_{ij} \equiv \partial_u C_{ij}.
\end{equation}
This is the quantity which characterizes the presence of radiation in these spacetimes (see, for example, Refs.~\cite{Geroch1977, Wald1999, Grant2021a}).

\subsection{GHP formalism} \label{sec:ghp}

The GHP formalism involves writing the metric in terms of a null tetrad $\{l^a, n^a, m^a, \bar m^a\}$, satisfying
\begin{equation}
  l^a n_a = -1, \qquad m^a \bar m_a = 1,
\end{equation}
with all other contractions vanishing.
The metric can then be written as
\begin{equation}
  g_{ab} = -2 l_{(a} n_{b)} + 2 m_{(a} \bar m_{b)}.
\end{equation}
The main quantities of interest are the \emph{Weyl scalars} $\Psi_i$, for $0 \leq i \leq 4$, defined by
\begin{subequations}
  \begin{align}
    \Psi_0 &\equiv C_{abcd} l^a m^b l^c m^d, \\
    \Psi_1 &\equiv C_{abcd} l^a n^b l^c m^d, \\
    \Psi_2 &\equiv C_{abcd} l^a m^b \bar m^c n^d, \\
    \Psi_3 &\equiv C_{abcd} l^a n^b \bar m^c n^d, \\
    \Psi_4 &\equiv C_{abcd} n^a \bar m^b n^c \bar m^d.
  \end{align}
\end{subequations}
The conventions for these quantities are taken from Ref.~\cite{Iozzo2020}.

In terms of Bondi coordinates, the tetrad that is used by the SpECTRE code's implementation of CCE is given by Eq.~(81) of Refs.~\cite{Moxon2020}:\footnote{To translate between the various quantities in this paper and in Ref.~\cite{Moxon2020} (in particular in their Eq.~10, where every quantity on the left-hand side has a $\mathring{{}}$), we have that
\begin{subequations}
  \begin{gather}
    \mathring \beta \equiv r \beta, \qquad \mathring W \equiv \frac{2V}{r^2}, \\
    \mathring U \equiv \sqrt{2} r^2 q_i \mc U^i, \quad \mathring Q \equiv \sqrt{2} r^2 e^{-2\beta/r} q^i \mc H_{ij} \frac{\partial}{\partial r} \left(\frac{\mc U^j}{r^2}\right), \\
    \mathring J \equiv \frac{1}{r} q^i q^j \mc C_{ij}, \qquad \mathring K \equiv \gamma,
  \end{gather}
\end{subequations}
where everything on the right-hand sides of these equations is defined using the conventions of this paper (relevant considering the differing definitions of $q^i$).}
\begin{subequations}
  \begin{align}
    l^a &\equiv \frac{1}{\sqrt 2} (\partial_r)^a, \\
    n^a &\equiv \sqrt{2} e^{-2 \beta/r} \bigg[(\partial_u)^a + \frac{1}{r^2} \mc U^i (\partial_i)^a \nonumber \\
    &\hspace{6.5em}- \frac{1}{2} \left(1 - \frac{2V}{r}\right) (\partial_r)^a\bigg], \\
    m^a &\equiv -\frac{1}{\sqrt{2} r} (\partial_i)^a q_j \left(\sqrt{1 + \gamma} h^{ij} - \frac{1}{r \sqrt{1 + \gamma}} \mc C^{ij}\right).
  \end{align}
\end{subequations}
This tetrad is defined in terms of a complex dyad $q^i$ on the sphere, defined (in spherical coordinates) by
\begin{equation}
  q^i \equiv -\frac{1}{\sqrt 2} \left[(\partial_\theta)^i + \frac{i}{\sin \theta} (\partial_\phi)^i\right].
\end{equation}
Note that this differs by a factor of $\sqrt 2$ from the conventions in Ref.~\cite{Moxon2020}.

This dyad is also used to construct the spin-raising and -lowering operators $\eth$ and $\bar \eth$, as follows: suppose that one has a tensor $v_{i_1 \cdots i_{p + q}}$ on the sphere.
One can then define a complex scalar $v$ by
\begin{equation}
  v \equiv v_{i_1 \cdot i_p j_1 \cdots j_q} q^{i_1} \cdots q^{i_p} \bar q^{j_1} \cdots q^{j_q};
\end{equation}
under a dyad rotation $q^i \to e^{i\psi} q^i$, $v \to e^{is \psi} v$, where $s = p - q$ is the \emph{spin weight} of $v$.
The operator $\eth$ is then defined by
\begin{equation}
  \eth v \equiv q^i q^{j_1} \cdots q^{j_p} \bar q^{k_1} \cdots \bar q^{k_q} \ms D_i v_{j_1 \cdots j_p k_1 \cdots k_q},
\end{equation}
while $\bar \eth$ is defined by $\bar \eth v = \overline{\eth \bar v}$.

Note that, if $v$ has spin weight $s$, then $\eth v$ has spin weight $s + 1$ and $\bar \eth v$ has spin weight $s - 1$.
As such, we can consider the eigenfunctions of $\eth \bar \eth$ (or, equivalently, $\bar \eth \eth$), which are the \emph{spin-weighted spherical harmonics} ${}_s Y_{\ell m}$.
In this paper, we will only need the following properties (using the conventions of the \textsc{scri} package~\cite{scri}\footnote{The \textsc{scri} package is a module for constructing, transforming, and working with waveform data at future null infinity~\cite{scri}.}):
\begin{subequations}
  \begin{align}
    \eth {}_s Y_{\ell m} &= \sqrt{\frac{(\ell - s) (\ell + s + 1)}{2}} {}_{s + 1} Y_{\ell m}, \\
    \bar \eth {}_s Y_{\ell m} &= -\sqrt{\frac{(\ell + s) (\ell - s + 1)}{2}} {}_{s - 1} Y_{\ell m},
  \end{align}
\end{subequations}
from which it follows that
\begin{subequations}
  \begin{align}
    \eth \bar \eth {}_s Y_{\ell m} &= -\frac{(\ell + s) (\ell - s + 1)}{2} {}_s Y_{\ell m}, \\
    \bar \eth \eth {}_s Y_{\ell m} &= -\frac{(\ell - s) (\ell + s + 1)}{2} {}_s Y_{\ell m}.
  \end{align}
\end{subequations}
As such, we can invert the action of $\eth$ or $\bar \eth$ on an individual spin-weighted spherical harmonic through a combination of applying $\bar \eth$ or $\eth$ (respectively) and dividing by an appropriate $\ell$- and $s$-dependent coefficient:
\begin{subequations} \label{eqn:swsh_inversion}
  \begin{align}
    \eth^{-1} {}_s Y_{\ell m} &= -\frac{2}{(\ell + s) (\ell - s + 1)} \bar \eth {}_s Y_{\ell m}, \\
    \bar \eth^{-1} {}_s Y_{\ell m} &= -\frac{2}{(\ell - s) (\ell + s + 1)} \eth {}_s Y_{\ell m}.
  \end{align}
\end{subequations}
Note that these equations \emph{only} provide an expression for $\eth^{-1}$ and $\bar \eth^{-1}$ acting on individual spin-weighted spherical harmonics.
As such, whenever we are inverting $\eth$ or $\bar \eth$, we will first expand on a basis of these functions.

Finally, this dyad can be used to define the \emph{strain} $h$~\cite{Mitman2020}:
\begin{equation}
  h \equiv r \bar q^i \bar q^j \mc C_{ij}.
\end{equation}
This quantity is important, since the leading order piece of $h$, together with the leading order pieces of the Weyl scalars $\Psi_i$, form the output of the CCE code.
In contrast, the quantities that are output by the \textsc{scri} package~\cite{scri,Boyle:2013nka,Boyle:2014ioa,Boyle:2015nqa}, which we use throughout the remainder of this paper, are $\psi_i$ and $\sigma$, which are defined in terms of these quantities by
\begin{gather}
  \psi_i \equiv \lim_{r \to \infty} r^{5 - i} (-\sqrt{2})^{2 - i} \Psi_i, \\
  \sigma \equiv \lim_{r \to \infty} \frac{\bar h}{2r}.
\end{gather}
Note that these could be defined as (coefficients in an expansion in $1/r$ of) the Weyl scalars (for $\psi_i$) and shear (for $\sigma$), defined as in Ref.~\cite{Geroch1973} with respect to \emph{some} tetrad, but this is not important for the discussion of this paper.

There are simple relationships between the $\psi_i$ and $\sigma$ and the functions appearing in the metric in Eq.~\eqref{eqn:bondi}.
These are given by
\begin{equation}
  \sigma = \frac{1}{2} q^i q^j C_{ij},
\end{equation}
and using Eq.~(94) of Ref.~\cite{Moxon2020}, we find that
\begin{subequations}
  \begin{align}
    \psi_0 &= -3 \left(\ord{0}{\mc E}_{ij} q^i q^j - 2 \sigma |\sigma|^2\right), \\
    \psi_1 &= -\frac{1}{2} q^i N_i, \\
    \psi_2 &= -m - \left(\Im[\eth^2 \bar \sigma] + \sigma \dot{\bar \sigma}\right),
  \end{align}
\end{subequations}
where dots denote partial derivatives with respect to $u$.

\subsection{Bianchi identities}

In Ref.~\cite{Grant2021b}, the evolution equations were given for $m$, $N_i$, and all of the $\ord{n}{\mc E}_{ij}$ (at least schematically, for $n > 1$).
These evolution equations arose through considering the vacuum Einstein equations.
In contrast, here, these same equations (for each of the $\psi_i$) are naturally derived from the differential Bianchi identity for $R^a{}_{bcd}$, which implies a differential equation for $C^a{}_{bcd}$ and therefore each of the $\Psi_i$, from which one can derive differential equations for each of the $\psi_i$.
These take the form of the following six equations (see, for example, the equations at the end of Sec.~9.8 of Ref.~\cite{Penrose1988b}; we are explicitly using the conventions here of Ref.~\cite{scri}):
\begin{subequations}
  \label{eqn:bianchiviolations}
  \begin{align}
    \dot \psi_0 &= \eth \psi_1 + 3 \sigma \psi_2, \label{eqn:bianchi_0} \\
    \dot \psi_1 &= \eth \psi_2 + 2 \sigma \psi_3, \label{eqn:bianchi_1} \\
    \dot \psi_2 &= \eth \psi_3 + \sigma \psi_4, \label{eqn:bianchi_2} \\
    \Im[\psi_2] &= -\Im[\eth^2 \bar \sigma + \sigma \dot{\bar \sigma}], \label{eqn:im_psi_2} \\
    \psi_3 &= -\eth \dot{\bar \sigma}, \label{eqn:psi_3} \\
    \psi_4 &= -\ddot{\bar \sigma}. \label{eqn:psi_4}
  \end{align}
\end{subequations}

It is also natural to write these equations in terms of a shifted version of $\psi_2$, defined in terms of the mass aspect by
\begin{equation}
  \tilde \psi_2 \equiv -m - i \Im[\eth^2 \bar \sigma] = \psi_2 + \sigma \dot{\bar \sigma},
\end{equation}
since then Eq.~\eqref{eqn:im_psi_2} becomes
\begin{equation}
  \Im[\tilde \psi_2] = -\Im[\eth^2 \bar \sigma],
\end{equation}
Combining Eqs.~\eqref{eqn:bianchi_2}, \eqref{eqn:psi_3}, and~\eqref{eqn:psi_4}, we find that
\begin{equation}
  \dot{\tilde \psi}_2 = -\eth^2 \dot{\bar \sigma} + \dot \sigma \dot{\bar \sigma}.
\end{equation}
Note that, as $\dot \sigma \dot{\bar \sigma}$ is purely real, the imaginary part of this equation contains no new information, and the real part is simply
\begin{equation} \label{eqn:mdot}
  \dot{m} = \Re[\eth^2 \dot{\bar \sigma}] - \dot \sigma \dot{\bar \sigma}.
\end{equation}

\subsection{Definition of our observables} \label{sec:moments}

In Ref.~\cite{Grant2021b}, it was shown that the so-called ``curve deviation observable'' of Ref.~\cite{Flanagan2019a}, an observable that a pair of observers could measure that was a natural generalization of the displacement memory effect, could be written in terms of (temporal) moments of the Bondi news.
Here, we adopt the following notation in terms of $\dot{\bar \sigma}$:
\begin{equation}
	\label{eq:momentofnews}
  \mc N_n (u_1, u_0; \tilde u) \equiv \frac{1}{n!} \int_{u_0}^{u_1} (\tilde u - u)^n \dot{\bar \sigma} (u) \ud u.
\end{equation}

Here, we explicitly keep the ``reference time'' $\tilde u$ unspecified.
In Ref.~\cite{Grant2021b}, these moments of the news could be related to pieces of the curve deviation observable if $\tilde u = u_0$.
We call such moments the ``Mellin moments'', since, for $u_0 = 0$ and $u_1 = \infty$, these moments are related to the Mellin transform $\mc M$ of $\dot{\bar \sigma}$, which is defined by
\begin{equation}
  \mc M_n \{f\} \equiv \int_0^\infty u^{n - 1} f(u) \ud u,
\end{equation}
for some arbitrary function $f$.
The exact relation is
\begin{equation}
  \mc N_n (\infty, 0; 0) = (-1)^n \mc M_{n + 1} \{\dot{\bar \sigma}\},
\end{equation}
where the factor of $(-1)^n$ was absorbed into the definition of $\mc N_n$ in Ref.~\cite{Grant2021b}.
This interpretation in terms of the Mellin transform arises in discussions of celestial holography~\cite{Freidel2022}.

However, as it is more convenient for this paper, we use the convention for the moments that $\tilde u = u_1$, and for brevity simply write
\begin{equation} \label{eqn:cauchy}
  \mc N_n (u_1, u_0) \equiv \mc N_n (u_1, u_0; u_1).
\end{equation}
These we call the ``Cauchy moments'', as they appear naturally in Cauchy's formula for multiple integration:
\begin{equation} \label{eqn:cauchy_formula}
  \begin{split}
    \mc N_n (u_1, u_0) &= \int_{u_0}^{u_1} \ud u_2 \mc N_{n - 1} (u_2, u_0) \\
    &= \int_{u_0}^{u_1} \ud u_2 \int_{u_0}^{u_2} \ud u_3 \cdots \int_{u_0}^{u_{n + 1}} \ud u\; \dot{\bar \sigma} (u).
  \end{split}
\end{equation}

In Sec.~\ref{sec:decomposition}, all results hold regardless of the chosen value of $\tilde u$, and so we adopt the ``generic'' definition of moments above, for an arbitrary $\tilde u$.
For our results in Sec.~\ref{sec:results}, however, we choose to use the Cauchy moments explicitly.

\section{``Charge'' and ``Flux'' Decomposition} \label{sec:decomposition}

In this section, we review the results of Ref.~\cite{Grant2021b}, which showed that one could write these moments of the news $\mc N_n (u_1, u_0; \tilde u)$ in terms of two contributions: a change in a ``charge'' and an integral of a ``flux''.
This is analogous to the split into ``linear'' and ``nonlinear'' memory of Ref.~\cite{Zeldovich1974} and Ref.~\cite{Christodoulou1991}, or of ``ordinary'' and ``null'' of Ref.~\cite{Bieri2013b}, in the case of vacuum general relativity.\footnote{In the case where there exist null matter fields, for example in Einstein-Maxwell theory, the stress-energy tensor of these fields is included in the ``null'' part defined by Ref.~\cite{Bieri2013b}, whereas here we would place it in a third category in this splitting.}

The general definition of a ``charge'' which we use here is that a charge is a quantity, such as the mass aspect $m$, which is constant in regions where the news vanishes.
In this sense, these are quasi-conserved quantities near null infinity.
There are a variety of prescriptions to define charges near null infinity, in the sense of quasi-conserved quantities which are conjugate to some symmetry~\cite{Wald1999, Compere2019, Grant2021a, Elhashash2021}; however, following Ref.~\cite{Grant2021b}, we merely provide a procedure by which one could construct these charges, and do not consider their relationship to (potential) symmetries at null infinity. Note, however, that our charges for the zeroth and first moments of the news are consistent with those in previous works, like~\cite{Wald1999,Compere2019}.

\subsection{First three moments of the news} \label{sec:moment_decomp}

Obtaining the zeroth moment of the news is the simplest case, as one can simply integrate Eq.~\eqref{eqn:mdot} once in time and rearrange terms to obtain:
\begin{equation} \label{eqn:disp}
  \Re\left[\eth^2 \mc N_0 (u_1, u_0; \tilde u)\right] = \Delta m(u_1, u_0) + \int_{u_0}^{u_1} \mc F_0 (u) \ud u,
\end{equation}
where
\begin{equation}
  \mc F_0 \equiv \dot \sigma \dot{\bar \sigma},
\end{equation}
and, for any function $f$,
\begin{equation}
  \Delta f(u_1, u_0) \equiv f(u_1) - f(u_0).
\end{equation}
Here Eq.~\eqref{eqn:disp} does not give the entire zeroth moment of the news: we are missing $\Im[\eth^2 \mc N_0 (u_1, u_0; \tilde u)]$, and there is no equation analogous to Eq.~\eqref{eqn:disp} from which this quantity can be derived.
For reasons discussed further in Refs.~\cite{Flanagan2015, Maedler2016}, it is expected to be fairly subdominant, which is consistent with Fig.~\ref{fig:M0hiearchy} and Fig.~\ref{fig:contributions}.\footnote{This is primarily due to the fact that magnetic memory is not sourced in binary black hole mergers, and is only known to appear for specific classes of spacetimes~\cite{Bieri:2020zki,Satishchandran2019}.} Last, note that Eq.~\eqref{eqn:disp} can also be written without the $\eth^{2}$ operator by simply inverting this operation via Eq.~\eqref{eqn:swsh_inversion}.

For the first moment, we instead consider equation~\eqref{eqn:bianchi_1}, which we write (in terms of $\tilde \psi_2$) as
\begin{equation}
  \dot \psi_1 = \eth \tilde \psi_2 - \dot{\bar \sigma} \eth \sigma - 3 \sigma \eth \dot{\bar \sigma}.
\end{equation}
Note that $\tilde \psi_2$ is non-zero even when $\dot \sigma = 0$.
As such, $\psi_1$ is \emph{not} a charge; to obtain a charge, we shift $\psi_1$ by a quantity coming from $\tilde \psi_2$:
\begin{equation}
  \tilde \psi_1 \equiv \psi_1 + (\tilde u - u) \eth \tilde \psi_2,
\end{equation}
where (for brevity) we do not explicitly denote the dependence of tilded quantities on $\tilde u$, and all quantities on the right-hand side are evaluated at $u$.
This quantity $\tilde \psi_1$ \emph{is} a charge, since
\begin{equation} \label{eqn:psi1tildedot}
  \dot{\tilde \psi}_1 = -(\tilde u - u) \eth^3 \dot{\bar \sigma} + (\tilde u - u) \eth \mc F_0 + \mc F_1,
\end{equation}
for
\begin{equation}
  \mc F_1 \equiv -\dot{\bar \sigma} \eth \sigma - 3 \sigma \eth \dot{\bar \sigma},
\end{equation}
which implies that $\dot{\tilde{\psi}}_{1}=0$ when $\dot{\sigma}=0$. Integrating as before, we find that
\begin{equation} \label{eqn:subdisp}
  \begin{split}
    \eth^3 \mc N_1 (u_1, u_0; \tilde u) &= -\Delta \tilde \psi_1 (u_1, u_0; \tilde u) + \int_{u_0}^{u_1} \mc F_1 (u) \ud u \\
    &\hspace{1.1em}+ \eth \int_{u_0}^{u_1} (\tilde u - u) \mc F_0 (u) \ud u,
  \end{split}
\end{equation}
where we write (for any function $\tilde f$ of $u$ and $\tilde u$)
\begin{equation}
  \Delta \tilde f (u_1, u_0; \tilde u) \equiv \tilde f (u_1, \tilde u) - \tilde f (u_0, \tilde u).
\end{equation}
On the right-hand side of Eq.~\eqref{eqn:subdisp} we have three terms instead of the two terms which appear in Eq.~\eqref{eqn:disp}.
By an integration by parts, however, it can be shown that the term involving $\mc F_0$ is (up to an integral) the same as the term involving $\mc F_0$ in Eq.~\eqref{eqn:disp}; as such, we do not consider it to be ``new'' information for determining the first moment of the news.

Finally, for the second moment, we consider equation~\eqref{eqn:bianchi_0}, which we write in terms of $\tilde \psi_2$ and $\tilde \psi_1$ as
\begin{equation}
  \dot{\psi}_0 = \eth \tilde \psi_1 + 3 \sigma \tilde \psi_2 - (\tilde u - u) \eth^2 \tilde \psi_2 - 3 \sigma^2 \dot{\bar \sigma}.
\end{equation}
Since none of the first three terms vanish when $\dot \sigma = 0$, $\psi_0$ is not, by itself, a charge; the procedure we applied above for defining $\tilde \psi_1$ shows that one should define the true charge $\tilde \psi_0$ by
\begin{equation} \label{eqn:tildepsi_0}
  \tilde \psi_0 \equiv \psi_0 + (\tilde u - u) (\eth \tilde \psi_1 + 3 \sigma \tilde \psi_2) - \frac{1}{2} (\tilde u - u)^2 \eth^2 \tilde \psi_2.
\end{equation}
Again, this quantity \emph{is} a charge, since
\begin{align}
  \dot{\tilde \psi}_1 - \frac{1}{2} (\tilde u - u) \eth \dot{\tilde \psi}_2 &= -\frac{1}{2} (\tilde u - u) \eth^3 \dot{\bar \sigma} \\\nonumber
  &\hspace{1.1em} + \frac{1}{2} (\tilde u - u) \eth \mc F_0 + \mc F_1,
\end{align}
which implies
\begin{equation} \label{eqn:psi0tildedot}
  \begin{split}
    \dot{\tilde \psi}_0 &= -\frac{1}{2} (\tilde u - u)^2 \eth^4 \dot{\bar \sigma} + \frac{1}{2} (\tilde u - u)^2 \eth^2 \mc F_0 \\
    &\hspace{1.1em}+ (\tilde u - u) (\eth \mc F_1 + \mc F^{\rm rad.}_{2, 1} + \mc F^{\rm nonrad.}_{2, 1}) + \mc F_{2, 0},
  \end{split}
\end{equation}
where
\begin{align}
  \mc F^{\rm nonrad.}_{2, 1} &\equiv -3 \frac{\ud}{\ud u} (m \sigma), \label{eqn:F_21_nonrad} \\
  \mc F^{\rm rad.}_{2, 1} &\equiv -3i \frac{\ud}{\ud u} (\sigma \Im[\eth^2 \bar \sigma]), \label{eqn:F_21_rad} \\
  \mc F_{2, 0} &\equiv -3 \sigma^2 \dot{\bar \sigma}.
\end{align}
This notation, with two subscripted numbers, is reminiscent of that in Eqs.~(4.29) and~(4.30) of Ref.~\cite{Grant2021b}.
Here, the first number indicates the moment of the news in question.
The second number reflects how the term arises in the equations of motion: if the second number is a zero (like for $\mc F_{2, 0}$), then it comes entirely from the evolution equation.
If the second number takes on some non-zero value $n$, then it indicates that there needed to be a correction to the function (in this case $\psi_0$) in order to make it a charge, and that the flux term involves $n$ derivatives with respect to $u$ of a term that came from the original evolution equation.
In the case of $\mc F_{2, 1}^{\rm rad./nonrad.}$, this can be seen in Eqs.~\eqref{eqn:F_21_nonrad} and~\eqref{eqn:F_21_rad}
The distinction between $\mc F^{\rm rad.}_{2, 1}$ and $\mc F^{\rm nonrad.}_{2, 1}$ is that the former contains only radiative data (that is, the shear $\sigma$), whereas the latter also contains some nonradiative data (in this case, the mass aspect $m$).

Finally, we integrate Eq.~\eqref{eqn:psi0tildedot} to obtain an equation for the second moment of the news:
\begin{widetext}
\begin{equation} \label{eqn:subsubdisp}
  \begin{split}
    \eth^4 \mc N_2 (u_1, u_0; \tilde u) &= -\Delta \tilde \psi_0 (u_0, u_1) + \frac{1}{2} \int_{u_0}^{u_1} (\tilde u - u)^2 \eth^2 \mc F_0 (u) \ud u + \int_{u_0}^{u_1} (\tilde u - u) \eth \mc F_1 (u) \ud u \\
    &\hspace{1.1em}+ \int_{u_0}^{u_1} \left\{(\tilde u - u) [\mc F^{\rm rad.}_{2, 1} (u) + \mc F^{\rm nonrad.}_{2, 1} (u)] + \mc F_{2, 0} (u)\right] \ud u.
  \end{split}
\end{equation}
\end{widetext}

\subsection{Ambiguities} \label{sec:ambiguities}

Note that there exist non-trivial modifications of the ``charges'' $m$, $\tilde \psi_1$, and $\tilde \psi_0$, that still satisfy the requirement that they are charges, namely that their time derivatives vanish in the absence of news.
By ``non-trivial'', we mean that the modifications do not vanish in regions where the news vanishes, which eliminates any modifications using $\psi_3$ or $\psi_4$, as they vanish in such regions.
In order for their time derivatives to vanish in the absence of news, any modifications must be constructed from quantities which themselves have that property, which eliminates all quantities other than $\tilde \psi_2$, $\tilde \psi_1$, $\tilde \psi_0$, and $\sigma$.

Next, note that any modifications need to have the correct dimensions, and the proper spin weights.
The quantities of interest have the following dimensions:
\begin{equation}
[\sigma] = [u]
\end{equation}
and
\begin{equation}
[\psi_i] = [u^{3 - i}].
\end{equation}
The spin weight of $\sigma$ is 2, and $\eth$ raises the spin weight by 1.
In addition, we have that action of complex conjugation flips the sign of the spin weight.
Finally, we have that $\psi_i$ has spin weight $2 - i$.

Finally, we note that we only allow quadratic and higher modifications---this follows from the fact that any linear modifications just change how the moments of $\dot \sigma$ are recovered.
Similarly, we do not allow any fractional or non-positive powers.
As such, for $m$, we see that there are no possible non-trivial modifications that can arise, as any possible modifications will necessarily have higher dimensions.
For $\tilde \psi_1$, the possible modifications can involve two factors of either $\sigma$, $\tilde \psi_2$, or a mixture of the two.
In order to get the correct spin weight, one needs to apply $\eth$ and $\bar \eth$ appropriately, or to use a $\bar \sigma$ instead of a $\sigma$.
Similarly, for $\tilde \psi_0$, the possible modifications involve either some cubic term involving $\sigma$ and $\tilde \psi_2$, or a quadratic term involving one factor of $\tilde \psi_1$ and one factor of either $\sigma$ or $\tilde \psi_2$.

The particular choice which we adopt for $\tilde \psi_1$ is as follows: consider instead
\begin{equation}
\hat \psi_1 \equiv \tilde \psi_1 + \frac{1}{2} (\bar \sigma \eth \sigma + 3 \sigma \eth \bar \sigma).
\end{equation}
This is the so-called ``Wald-Zoupas'' choice for this charge~\cite{Wald1999, Flanagan2015, Grant2021a}.
From Eq.~\eqref{eqn:psi1tildedot}, this quantity has the property that
\begin{equation}
  \dot{\hat \psi}_1 = -(\tilde u - u) \eth^3 \dot{\bar \sigma} + (\tilde u - u) \eth \mc F_0 + \hat{\mc F}_1,
\end{equation}
where
\begin{equation}
  \hat{\mc F}_1 = \frac{1}{2} \left(\bar \sigma \eth \dot \sigma + 3 \dot \sigma \eth \bar \sigma\right) - \interchange{\sigma}{\dot \sigma}.
\end{equation}
As in Eq.~\eqref{eqn:subdisp}, we can use this ``modified'' definition in order to write down an equation for the first moment of the news: the only change is that $\tilde \psi_1$ is replaced with $\hat \psi_1$ and $\mc F_1$ is replaced with $\hat{\mc F}_1$.
Finally, when we modify $\tilde \psi_1$ into $\hat \psi_1$, it makes sense to also modify $\tilde \psi_0$ into $\hat \psi_0$, which is defined as in Eq.~\eqref{eqn:tildepsi_0}, but with $\hat \psi_1$ instead of $\tilde \psi_1$.
Once this modification is performed, it follows that Eq.~\eqref{eqn:subsubdisp} can be written in the same form, but with $\tilde \psi_0$ replaced with $\hat \psi_0$ and $\mc F_1$ replaced by $\hat{\mc F}_1$: there are no modifications to the other flux terms.

To allow for easier comparison to the literature, we adopt the Wald-Zoupas convention for the charges and fluxes for the remainder of the paper.
Note, however, that only the sum of the charge and flux contributions is invariant, and so any claims about the relative magnitude the contributions, or their morphology, will be affected by this choice.
In particular, the particular shape of the flux contribution to the center-of-mass memory, as given in the left panel of Fig.~\ref{fig:structure}, will be different in the case where one makes another choice, although the discrepancy appears to be small and the qualitative features the same.

\subsection{The shear contributing to a moment}

\begin{figure*}[t]
	\centering
	\includegraphics[width=0.95\textwidth]{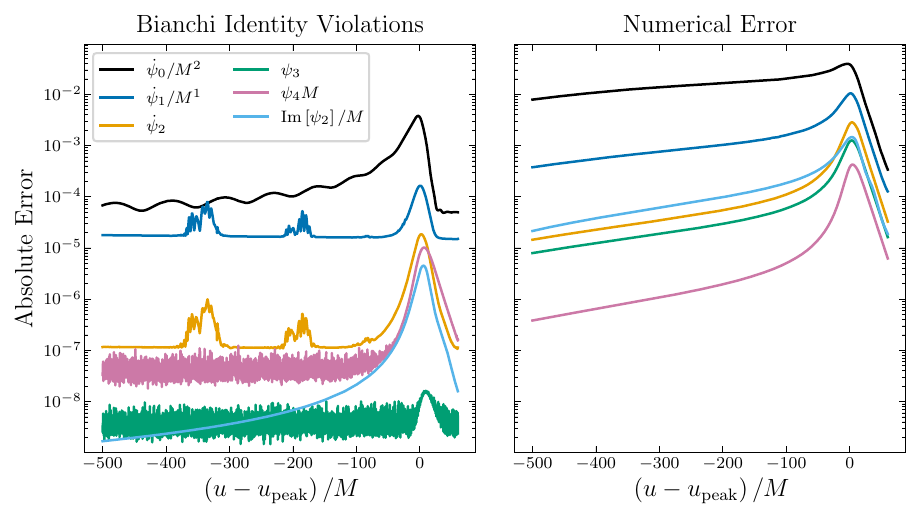}
	\caption{The absolute error of the Bianchi identity violations (left) and a conservative estimate of the numerical error (right).
		The violation of the Bianchi identities is computed by taking the difference of the terms on the left- and right-hand sides of Eqs.~\eqref{eqn:bianchiviolations}, while the numerical error is computed by taking the difference of the two highest-resolution simulations.
	}
	\label{fig:bianchiviolations}
\end{figure*}

So far, we have considered Eqs.~\eqref{eqn:disp}, \eqref{eqn:subdisp}, and~\eqref{eqn:subsubdisp} to be equations for the moments of the news in terms of $\psi_0$, $\psi_1$, $\psi_2$, and $\sigma$.
However, they are best thought of as \emph{consistency conditions} that must be satisfied by $\sigma$ directly, as it appears on both sides of the equation (as it appears in the definition of the moments of the news).
In the case where one could determine the shear from the moment, one could then ask the following question: in each of these equations, which parts of the shear come from which terms on the right-hand side?
These are what we call the ``shear contributing to a moment''.

The simplest relation between the moment of the news $\mc N_n$ and the shear $\sigma$ is given by the case where we use the ``Cauchy moments'' defined above in Eq.~\eqref{eqn:cauchy}.
From Eq.~\eqref{eqn:cauchy_formula}, it is apparent that
\begin{equation}
  \mc N_{n - 1} (u_1, u_0) = \partial_{u_1} \mc N_n (u_1, u_0).
\end{equation}
Finally, we have that
\begin{equation}
  \bar \sigma(u_1) - \bar \sigma(u_0) = \mc N_0 (u_1, u_0);
\end{equation}
note that, since the moments of the news depend on the derivative of $\sigma$, one cannot recover the \emph{entire} $\sigma$.
As such, we simply define the ``shear contribution'' from each term by the $n$th derivative of the corresponding term in the moment of the news.

We now consider each moment: for the zeroth moment of the news, the charge and flux contributions are the change in the mass aspect $\Delta m$ and the integral of the flux $\mc F_0$:
\begin{equation} \label{eqn:shear_0}
  \begin{split}
   \Re\{\eth^2 [\bar \sigma(u_1) - \bar \sigma(u_0)]\} &= \Delta m(u_1, u_0) \\
   &\hspace{1.1em} + \int_{u_0}^{u_1} \mc F_0 (u) \ud u.
  \end{split}
\end{equation}
In contrast, for the first moment, the charge contribution to the shear is the derivative of $\Delta \hat \psi_1$, and the flux contribution is $\hat{\mc F}_1$ (\emph{not} its integral):
\begin{equation} \label{eqn:shear_1}
  \begin{split}
    \eth^3 [\bar \sigma(u_1) - \bar \sigma(u_0)] &= -\frac{\partial}{\partial u_1} \Delta \hat \psi_1 (u_1, u_0; u_1) + \hat{\mc F}_1 (u_1) \\
    &\hspace{1.1em}+ \eth \int_{u_0}^{u_1} \mc F_0 (u) \ud u.
  \end{split}
\end{equation}
For the second moment, the charge contribution to the shear is the second derivative of $\Delta \hat \psi_0$, and the flux contributions are $\mc F^{\rm rad./nonrad.}_{2, 1}$ and $\dot{\mc F}_{2, 0}$:
\begin{equation} \label{eqn:shear_2}
  \begin{split}
    \eth^4 [\bar \sigma(u_1) - \bar \sigma(u_0)] &= -\frac{\partial^2}{\partial u_1^2} \Delta \hat \psi_0 (u_1, u_0; u_1) + \dot{\mc F}_{2, 0} (u_1) \\
    &\hspace{1.1em}+ \mc F_{2, 1}^{\rm rad.} (u_1) + \mc F_{2, 1}^{\rm nonrad.} (u_1) \\
    &\hspace{1.1em}+ \eth \hat{\mc F}_1 (u_1) + \eth^2 \int_{u_0}^{u_1} \mc F_0 (u) \ud u.
  \end{split}
\end{equation}
These are the quantities which we plot in Sec.~\ref{sec:results}.

Note that, in order to determine the contributions to the shear, one must invert the angular operators that act on the left-hand sides of Eqs.~\eqref{eqn:disp}, \eqref{eqn:subdisp}, and~\eqref{eqn:subsubdisp}, which can be easily done mode-by-mode on a basis of spin-weighted spherical harmonics using Eq.~\eqref{eqn:swsh_inversion}.

\section{Results} \label{sec:results}

\begin{figure*}[t]
	\centering
	\includegraphics[width=0.95\textwidth]{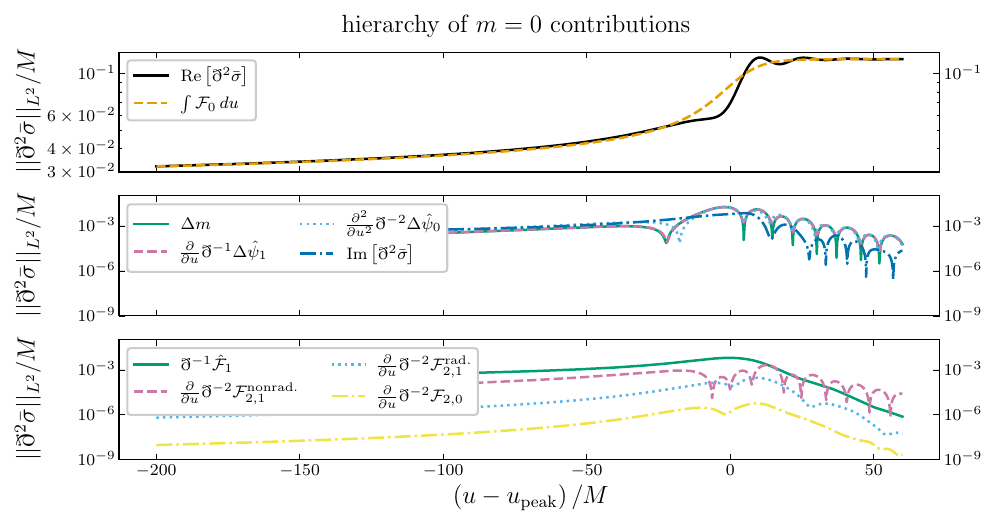}
	\caption{The $L^2$ norm of the $m=0$ modes of each charge and flux term's contribution to $\eth^2 \bar \sigma$, the spin-weight $0$ version of the shear.
    The curves are sorted by the magnitude of the time integral of their $L^2$ norm, that is, Eq.~\eqref{eq:norm}, with the largest being first.
    We also include the real and imaginary parts of $\eth^2 \bar \sigma$ for comparison.
	}
	\label{fig:M0hiearchy}
\end{figure*}

\subsection{Numerical Data}

For our numerical analysis of the moments of the news, we examine a simulation of an equal-mass, non-spinning binary black hole system.
This simulation comes from the SXS catalog and is identified by the ID \texttt{SXS:BBH\_ExtCCE:0001}~\cite{Boyle:2019kee}.
Because we are interested in not only the strain, but also the Weyl scalars at future null infinity, to perform our computations we use the waveforms that are produced by SpECTRE's CCE module~\cite{Moxon2020,Moxon:2021gbv,spectrecode}.
Last, before using the waveform and the Weyl scalars that are output by CCE, we first map this asymptotic waveform data to the post-Newtonian (PN) BMS frame using a 3PN waveform, the routine outlined in Ref.~\cite{Mitman:2022kwt}, and the python code \textsc{scri}~\cite{scri,Boyle:2013nka,Boyle:2014ioa,Boyle:2015nqa}.

To illustrate the success of CCE at extracting the asymptotic data for this system, in Fig.~\ref{fig:bianchiviolations} we show the Bianchi identity violations and an estimate of the numerical error for the asymptotic data of this system.
More specifically, in the left panel we plot the $L^2$ norm of the difference in the right- and left-hand sides of Eqs.~\eqref{eqn:bianchiviolations}.
Meanwhile, in the right panel we plot the $L^2$ norm of the difference of the two highest-resolution simulations, after we map the lower-resolution simulation to the BMS frame of the higher-resolution simulation~\cite{Mitman:2022kwt}.
We define the $L^{2}$ norm of a function $f$ (over the two-sphere) as
\begin{equation}
||f||_{L^{2}}=\sqrt{\int |f|^{2}\,d\Omega}.
\end{equation}

We stress  that the numerical error estimate that we provide is more an error estimate on the lower-resolution simulation, and that the errors for the simulation that we examine will be smaller than what is shown.
With this in mind, we find that the violation of the Bianchi identities and the numerical error for the Weyl scalars $\psi_2$, $\psi_3$, and $\psi_4$ are particularly low with absolute errors of $\lesssim 0.001\%$, whereas for the Weyl scalars $\psi_0$ and $\psi_1$ the Bianchi violations are noticeably larger.
We attribute this to the fact that both $\psi_0$ and $\psi_1$ are more influenced by backscattering physics, that is, the fact that information should be traveling back and forth between the Cauchy evolution and CCE, than the other Weyl scalars.
In fact, this notion has even been confirmed by the recent implementation of SpECTRE's Cauchy-characteristic matching code, which shows that the violation of the $\psi_0$ and $\psi_1$ Bianchi identities is much smaller than what CCE usually yields, at least for problem of evolving a Teukolsky wave~\cite{Ma:2023qjn}.\footnote{In particular, see Fig. 7 of Ref.~\cite{Ma:2023qjn}.}
Nonetheless, because the violations of the $\psi_0$ and $\psi_1$ Bianchi identities still correspond to absolute errors of $\lesssim 1\%$, the results of Fig.~\ref{fig:bianchiviolations} indicate that the errors on our asymptotic data are reasonable enough to perform the following analysis.
Furthermore, we find that the results drawn from the following analysis are unchanged if one uses the lower-resolution simulation instead of the higher-resolution simulation.

\subsection{Evaluation of Charges and Fluxes}

\begin{figure*}[t]
	\centering
	\includegraphics[width=0.95\textwidth]{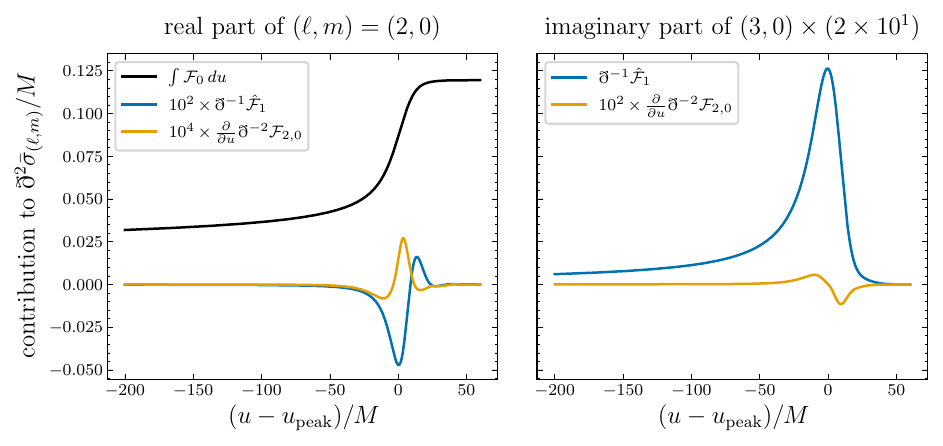}
	\caption{The flux term contributions to the real part of the $(2,0)$ mode (left) and the imaginary part of the $(3,0)$ mode (right), which represent the dominant modes for the displacement and spin memories.
    In these plots, we scale the $(3,0)$ mode as well as the individual flux terms by varying amounts to make their shapes easier to compare.
	}
	\label{fig:structure}
\end{figure*}

\begin{figure*}[t]
	\centering
	\includegraphics[width=0.95\textwidth]{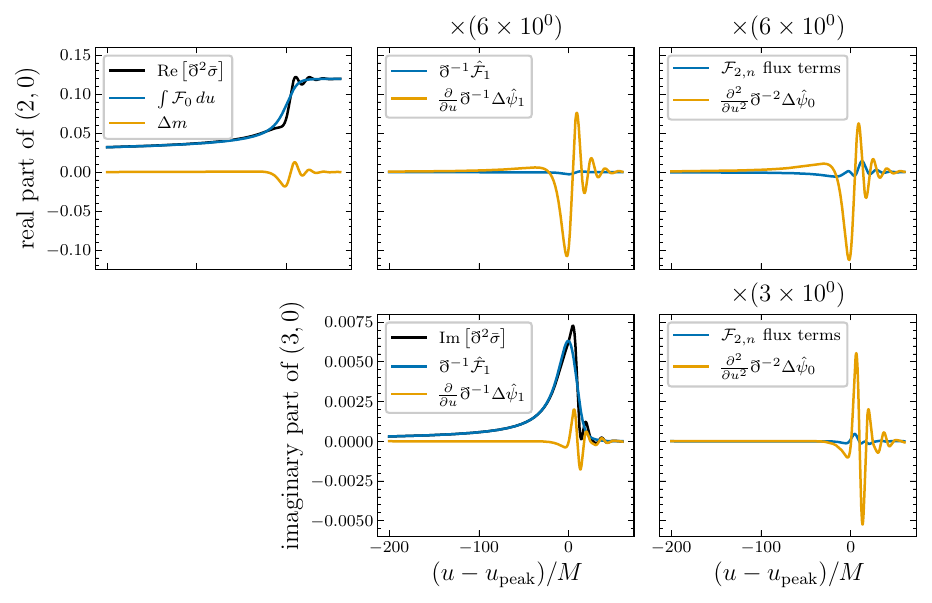}
	\caption{Contribution from the fluxes and the charges to the real part of the $(2,0)$ mode (top row) and the imaginary part of the $(3,0)$ mode (bottom row) of $\eth^{2}\bar{\sigma}$. Each column corresponds to the contribution from terms appearing in that moment of the news, e.g., the first column shows terms that appear in the balance law for the zeroth moment of the news. Certain panels are scaled by certain factors to make the comparison of each term more straightforward. For the second moment of the news, we show the sum of the $\mathcal{F}_{2,n}$ fluxes, for $n\in\{0,1\}$, since the individual fluxes are notably small.}
	\label{fig:contributions}
\end{figure*}

We now illustrate the hierarchy of the various charge and flux terms in terms of their contribution to the shear for our binary black hole simulation.
In Fig.~\ref{fig:M0hiearchy} we show the $L^2$ norm of the $m=0$ modes of each term's contribution to $\eth^{2}\bar{\sigma}$.
We organize the curves that are shown in descending order with respect to the magnitude of the retarded time integral of their $L^2$ norms:
\begin{equation} \label{eq:norm}
	\sqrt{\int ||\cdot||_{L^{2}}^{2} \ud u}.
\end{equation}
As can be seen, $\mathcal{F}_0$ is the largest contribution to the shear, followed by the various charge terms, and then the other flux terms, with $\hat{\mathcal{F}}_{1}$ being the second largest flux term.
Note that we restrict our analysis of these terms to the $m=0$ modes because these modes are the modes which typically are the most non-oscillatory~\cite{Mitman2020}.
Therefore, by focusing on these modes we have a better understanding of the hierarchy of the non-oscillatory contributions to the shear.

With the hierarchy of these terms understood, we now turn to examining the contributions of the flux terms to the primary non-oscillatory modes: the real part of the $(2,0)$ mode and the imaginary part of the $(3,0)$ mode.
We focus on the flux terms because, for binary systems, these terms should contribute more to net changes in the various moments of the news than the charge terms, which will dominate for unbound systems.
In Fig.~\ref{fig:structure} we show the contributions from $\mathcal{F}_0$, $\hat{\mathcal{F}}_1$, and $\mathcal{F}_{2, 0}$ to the real part of the $(2,0)$ mode and the imaginary part of the $(3,0)$ mode of the shear.
As can be seen, the $\mathcal{F}_0$, $\mathrm{Im}[\hat{\mathcal{F}}_1]$, and $\mathrm{Im}[\mathcal F_{2, 0}]$ curves, i.e., the curves corresponding to the displacement, spin, and magnetic-parity first higher memories, exhibit morphologies that one would expect based on the moment of the news that they appear in: a step-like function, a delta-like function, and the derivative of a delta-like function.
The $\mathrm{Re}[\hat{\mathcal{F}}_1]$ and the $\mathrm{Re}[\mathcal{F}_{2, 0}]$ curves, i.e., the curves corresponding to the center-of-mass and the electric-parity first higher memory, however, instead appear to break from this ``derivatives of step functions''-like behavior.
We believe that the reason these terms do not exhibit the expected behavior is because of one of the following reasons. First, it could be that one simply needs to perform a correction---like the Wald-Zoupas correction discussed in Sec.~\ref{sec:ambiguities}.
Or, and perhaps more likely, it may be that there are quasi-normal mode (QNM) oscillations that are obscuring the morphology.
In particular, because these flux terms are second- and third-order combinations of the shear, there could be nonlinear combinations of co-rotating and counter-rotating QNMs that yield a nontrivial, oscillatory contribution to these terms which would otherwise look like derivatives of step functions~\cite{Mitman:2022qdl,Cheung:2022rbm}.\footnote{We thank David Nichols for bringing this observation to our attention.}

Finally, in Fig.~\ref{fig:contributions} we show how the nonradiative fluxes and the charges contribute to these modes.
As is shown, the main observation is that beyond the curves that correspond to the displacement and the spin memories, which exhibit dominant contributions from the fluxes, the remaining curves are dominated by contributions from the charges.
This is expected because of the following heuristic argument.

First, note that the total signal in Fig.~\ref{fig:contributions} is predominantly non-oscillatory, for both the electric and magnetic parts of the shear.
Flux contributions, similarly, seem to mostly be non-oscillatory, whereas charge contributions appear to mostly be oscillatory.
By comparing Eqs.~\eqref{eqn:shear_0}, \eqref{eqn:shear_1}, and~\eqref{eqn:shear_2}, it is apparent that (up to angular operators) the contribution to the shear from the order $n$ charge can be written as the sum of the order $n + 1$ charge and flux contributions (for a similar discussion, see Ref.~\cite{Siddhant2024}).
Except for the case when these two contributions sum to the total (mostly non-oscillatory) signal, they are therefore summing to a piece which is mostly oscillatory.
As such, the mostly oscillatory charge contribution of order $n + 1$ must be dominant over the order $n + 1$ flux contribution, and this pattern should persist for all of the higher memory effects.

\section{Discussion} \label{sec:discussion}

In this work we provide straightforward expressions for the first three moments of the news (that is, the displacement, drift, and ballistic memories) in terms of charges and radiative and non-radiative fluxes as well as the first numerical calculation of these terms in the context of a binary black hole merger.
Besides providing a hierarchy for these terms in terms of how much they contribute to the various memory effects, we also explore how much the morphology of these contributions match ``derivatives of step functions'', which is what they should resemble if one assumes that they predominantly source non-oscillatory memory effects.
In particular, we find that while the $\mathcal{F}_0$ and $\mathcal{F}_1$ flux contributions to the displacement and spin memory effects resemble step- and delta-like functions, the other fluxes break from this expected behavior.
A possible explanation for this, however, could simply be that to make these terms exhibit this behavior, then one must perform a correction like the Wald-Zoupas correction that is used for the first moment of the news~\cite{Wald1999, Flanagan2015, Grant2021a} or remove the nontrivial QNM contributions that are sourced during the ringdown phase~\cite{Mitman:2022qdl,Cheung:2022rbm}.

Apart from this, we also find that these higher memory contributions are particularly small, relative to their displacement and spin memory counterparts.
Specifically, the contributions to the second moment of the news tend to be roughly two orders of magnitude smaller than the contributions to the zeroth moment of the news.
Consequently, while these terms are interesting from a theoretical perspective because they represent certain nonlinear features of Einstein's equations, this means that there is little hope to observe these effects until future detectors, like LISA, come online.

\acknowledgments

We thank David Nichols and Siddhant Siddhant for comments on an earlier version of this manuscript.
Computations for this work were performed with the Wheeler cluster at Caltech.
K.M. acknowledges the support of the Sherman Fairchild Foundation and NSF Grants No. PHY-2011968, PHY-2011961, PHY-2309211, PHY-2309231, OAC-2209656 at Caltech.
A.M.G. acknowledges the support of the Royal Society under grant \mbox{number RF\textbackslash ERE\textbackslash 221005}.

\bibliography{refs}

\end{document}